# *Ab-inito* study of low temperature magnetic properties of double perovskite $Sr_2FeOsO_6$

Sudipta Kanungo[1], Binghai Yan[1,2*], Martin Jansen[1,3], and Claudia Felser[1,4]

[1] *Max -Planck - Institut für Chemische Physik fester Stoffe, 01187 Dresden, Germany*
[2] *Max-Planck-Institut für Physik komplexer Systeme, 01187, Dresden, Germany*
[3] *Max-Planck-Institut für Festkörperforschung, 70569 Stuttgart, Germany*
[4] *Johannes Gutenberg-Universit, Institut für Anorganische Chemie und Analytische Chemie, 55128 Mainz, Germany*

*Email: yan@cpfs.mpg.de

Using density-functional theory calculations, we investigated the electronic structure and magnetic exchange interactions of the ordered 3*d*-5*d* double perovskite $Sr_2FeOsO_6$, which has recently drawn attention for interesting antiferromagnetic transitions. Our study reveals the vital role played by long-range magnetic exchange interactions in this compound. The competition between the ferromagnetic nearest neighbor Os-O-Fe interaction and antiferromagnetic next nearest neighbor Os-O-Fe-O-Os interaction induces strong frustration in this system, which explains the lattice distortion and magnetic phase transitions observed in experiments.

## I. INTRODUCTION

Double-perovskite oxides, $A_2BB'O_6$, are perovskite-derived systems with considerable flexibility to exhibit rich magnetic properties owing to the coexistence of two magnetic sites B and B'. The composition and nature of B and B' cations drastically affect the magnetic properties of double perovskites, which can result in metallicity (e.g $Sr_2CrReO_6$ [ref. 1]), half-metallicity (e.g $A_2CrWO_6$ [ref. 2]), ferromagnetism (e.g $Ca_2CrSbO_6$ [ref. 3], $La_2CoMnO_6$ [ref. 4]), colossal magnetoresistance ($Sr_2FeMoO_6$ [ref. 5]), multiferroicity (e.g $Ba_2NiMnO_6$ [ref. 6]), magneto-optic properties (e.g $Sr_2CrOsO_6$, $Sr_2CrReO_6$ [ref. 7]) and magneto-dielectric properties (e.g $La_2NiMnO_6$ [ref. 8]). In this context, 3*d*-5*d* combinations at B-B' sites are quite interesting. It is generally believed that the 3*d* element exhibits more electron correlation than its 5*d* counterparts. Therefore, it is expected that the B and B' sub-lattices behave differently. The most extensively studied 3*d*-5*d* double perovskites are $A_2BWO_6$ (refs. 9-11) and $A_2BReO_6$ (refs. 12,13) (where B represents a 3*d* element). For example, $Sr_2BOsO_6$, (B= Mn, Fe, Ni and Cu) compounds form a tetragonal structure, whereas $Sr_2VOsO_6$ and $Sr_2GaOsO_6$ systems crystallize into the monoclinic and cubic lattices, respectively.[14-17] Moreover, $Sr_2CrOsO_6$ (ref. 18) and $Sr_2CoOsO_6$ (refs. 19,20) undergo structural phase transitions from cubic to rhombohedral phases and tetragonal to monoclinic phases, respectively.

Recently, the structural and magnetic properties of $Sr_2FeOsO_6$ ordered double perovskite were experimentally investigated. Magnetic susceptibility and specific heat measurements show two magnetic transitions: a paramagnetic to antiferromagnetic (AFM) transition at 140 K and an AFM to AFM transition at 67 K.[20] Neutron diffraction[21] reveals that these two AFM phases differ only in the spin ordering along the c axis of the lattice, the Fe-Os-Fe-Os chain along the c axis exhibits a ferromagnetic (FM) sequence (↑↑↑↑ ) in the first AFM phase (labeled AF1 with $T_N$ = 140 K) and a mixed sequence (↑↑↓↓) in the second AFM phase (AF2 with $T_N$ = 67 K ). Previous density functional theory (DFT) calculations and experimental measurements indicated that the new spin sequence in the AF2 phase co-exists with alternate elongation and contraction of Fe-Os distances along the *c*-axis. The subtle balance between the effect of strong frustration and structural deformation was speculated to play a crucial role in the AF1-AF2 transition. However, a thorough understanding of the correlation between lattice distortion and magnetic exchange coupling has not yet been addressed.

In this study, we investigate the electronic and magnetic properties of $Sr_2FeOsO_6$ by *ab-initio* calculations. The magnetic exchange interactions were estimated from total energy calculations based on an effective Ising model.[22] Along the *c*-axis, the strong competition between short-range Os-O-Fe super-exchange coupling and long range Os-O-Fe-O-Os coupling induces the lattice distortion in the AF2 phase, and this distortion is due to strong frustration. In addition, exchange coupling for the Os-Os site is found to be much larger than the corresponding Fe-Fe coupling, which is due to the fact that Os-5$d$ states are much more extended than Fe-3$d$ states.

## II. CALCULATION METHOD AND CRYSTAL STRUCTURE

The density functional theory (DFT) calculations were performed within the plane-wave basis set based on a pseudopotential framework as implemented in the *Vienna ab-initio simulation package* (VASP).[23] The generalized gradient approximation (GGA) was employed following the Perdew-Burke-Ernzerhof prescription.[24] The missing correlation effect beyond GGA was taken into account through GGA + U calculations.[25,26] For the plane-wave basis, a 600 eV plane-wave cutoff was applied. A k-point mesh of 8 × 8 × 6 in the Brillouin zone was used for self-consistent calculations. The construction of the low-energy Hamiltonian in the *ab-initio* derived Wannier function basis[27] has been achieved through the downfolding technique. Starting from a full DFT calculation, the down-folding technique, involves the adoption of Fe-3$d$ and Os-5$d$ states for the projection. $Sr_2FeOsO_6$ crystallizes in a tetragonal crystal structure with space group I4/m.[20] This structure consists of alternating corner sharing $FeO_6$ and $OsO_6$ octahedra, along all directions. The Sr atoms are situated at the void positions between two types of octahedra. Both, $FeO_6$ and $OsO_6$, are distorted from a perfect octahedral structure. Along the *c*-axis, $FeO_6$ is elongated, while $OsO_6$ is compressed. The Fe-O-Os-O-Fe-O-Os chain is a perfectly straight chain (∠Fe-O-Os=180º) along the crystallographic *c*-axis, whereas the chain exhibits a zig-zag shape (∠Fe-O-Os=177.6º) along the crystallographic a and b axes. For DFT calculations, we adopt the experimental crystal structure of tetragonal lattice that corresponds to the structure measured at 78 K.[20]

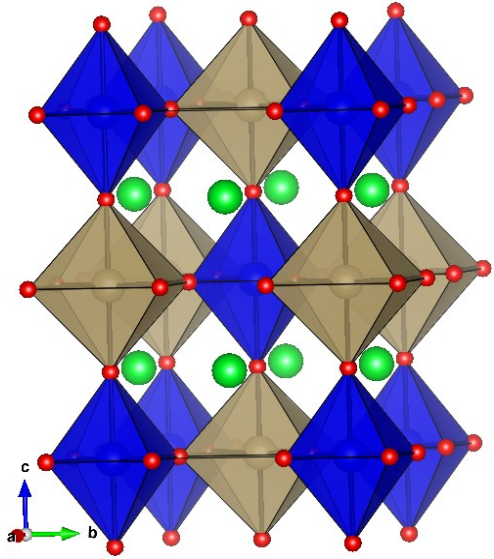

*Fig.1: Crystal structure of $Sr_2FeOsO_6$ in the tetragonal phase. Blue and brown octahedra represent $FeO_6$ and $OsO_6$ octahedra, respectively. Green and red balls represent Sr and O atoms, respectively.*

## III. RESULTS AND DISCUSSIONS

The calculated density of states (DOS) is shown in Fig.2. The Sr states lie far above the Fermi level ($E_f$), and this finding is consistent with the nominal $Sr^{+2}$ valence state. Both, Fe-3$d$ and Os-5$d$ states, contribute to the DOS at $E_f$ and Fe and Os $d$ states show significant mixing with the O-2$p$ states. Note that in the absence of spin polarization, the band structure shows a metallic character, while the spin polarized calculations within the GGA, assuming small Coulomb correlation at the 3$d$ (Fe) and 5$d$ (Os) sites, drive the insulating solution. We note that calculated magnetic moments depend sensitively on spin-orbit coupling (SOC) and effective Coulomb U. For example, GGA+SOC (GGA) gives $\mu_{Fe}$=2.8 (2.8) $\mu_B$ and $\mu_{Os}$=0.9 (1.7) $\mu_B$; GGA+U+SOC (GGA+U) shows $\mu_{Fe}$=3.9 (3.8) $\mu_B$ and $\mu_{Os}$=1.6 (2.5) $\mu_B$, where $U^{Fe}_{eff}$ = 4 eV and $U^{Os}_{eff}$ = 2 eV. We show the DOS of the GGA+U calculation in Fig. 2. The Fe-3d states are completely filled in one spin channel and are completely empty in the other spin channel. Along with the calculated magnetic moment, these findings suggest that, Fe is in the +3 ($3d^5$) valence electronic state with a high spin configuration resulting in S = 5/2, and Os is in the +5 ($5d^3$) valence electronic state with S = 3/2. In the octahedral environment, $d$ states of Fe and Os represent $t_{2g}$-$e_g$ type energy splitting. Due to the presence of tetragonal distortion, $t_{2g}$ and $e_g$ states are no longer degenerate. Elongation of FeO$_6$ octahedra along the c- axis splits the triply degenerate $t_{2g}$ states into the lower, almost degenerate $d_{yz}$-$d_{zx}$ bands ($e^\pi_g$ symmetry) and upper $d_{xy}$ band ($a_{1g}$ symmetry). The doubly degenerate $e_g$ states split into lower $d_{z2}$ and higher $d_{x2-y2}$ bands. Since the distortions of OsO$_6$ octahedra are in the opposite direction, *i.e.* the OsO$_6$ octahedra are contracted along the crystallographic *c*-axis, the level splitting also shows an opposite trend. Compared to Fe-*d* states, the band width of Os-*d* states is much wider. Therefore the $t_{2g}$-$e_g$ crystal field splitting gap of Os-5*d* states is much larger than the splitting gap of Fe-3*d* states. Consequently, the effective spin model for this compound can be constructed in terms of the Fe-*d* ($t_{2g}$-$e_g$) and Os-$t_{2g}$ degrees of freedom.

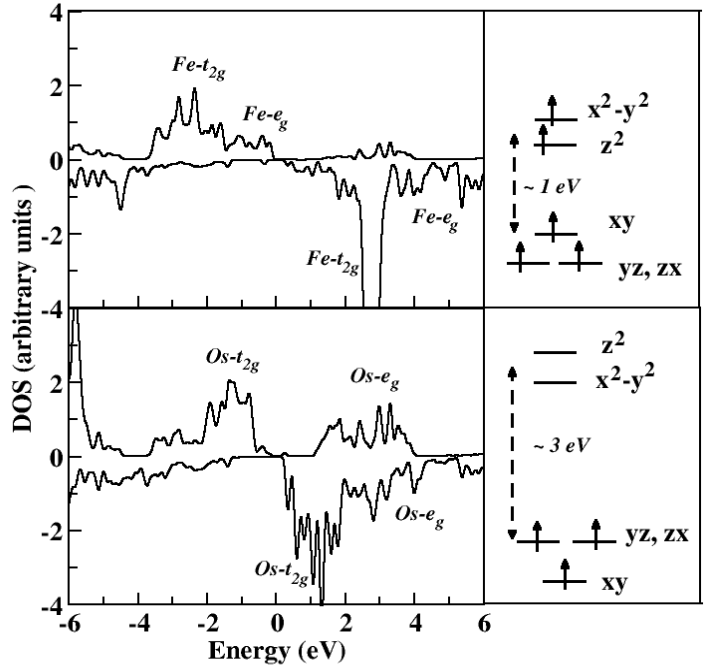

*Fig.2. GGA spin polarized density of states. The upper and lower panels at the left show the projected DOS for Fe-3d and Os-5d states, respectively. The two channels for each panel represent majority and minority spin channels. The Fermi level is marked at zero on the energy scale. The upper and lower panels at the right show the energy level splitting and the occupancies for Fe-3d and Os-5d states, respectively, obtained from DFT calculations.*

Magnetic exchange interactions can be evaluated based on the understanding of hopping integrals and on-site energies. Using the extended Kugel-Khomskii model, the expression for the AFM and FM super-exchange interactions are given by $J^{AFM} \propto \sum_{ij} \frac{t_{ij}^2}{U+\Delta_{ij}}$ and $J^{FM} \propto -\sum_{ij} \frac{t_{ij}^2 J_H}{(U+\Delta_{ij})(U+\Delta_{ij}-J_H)}$ respectively,[28,29,30] where $t_{ij}$ is the hopping interaction connecting the i$^{th}$ and j$^{th}$ sites, U is the on-site Coulomb interaction i.e the energy cost in putting two electrons at the same site, $\Delta_{ij}$ is the difference in energy level positions between i$^{th}$ and j$^{th}$ levels, and $J_H$ is the Hund's coupling. Adopting this approach requires the correct choice of Hubbard on-site correlation U, of the Hund's exchange $J_H$ and of a proper estimate of charge transfer energy $\Delta$ for different orbitals. However, because of the complicated super-exchange paths involving different types of atoms, such energies are difficult to estimate. Therefore, we adopt an alternative route for evaluating the effective magnetic exchange interactions. Technically, magnetic exchange is the difference between the energy costs corresponding to the FM and AFM spin arrangements. We therefore attempted to estimate the magnetic exchange interactions using the total energy calculations of various spin configurations, and we then mapped the DFT total energy to the corresponding Ising Model[22] with the form $E^{total} = \sum_{ij} J_{ij} \vec{S}_i \cdot \vec{S}_j$ , where $J_{ij}$ is the exchange interaction between the $i^{th}$ and $j^{th}$ sites and $S_i$ and $S_j$ are the effective spin values at the $i^{th}$ and $j^{th}$ sites, respectively. Although such a scheme has several difficulties, such as the choice of spin configurations, choice of basis sets, and exchange-correlation functional, this method is nonetheless successful in providing an indicative estimate. Since there is no full proof way to estimate the correct U values for the *d* states, we prefer the GGA functional without any artificial U values to estimate the exchange coupling. In order to understand the effect of magnetic frustration, we adopt the tetragonal lattice that corresponds to the structure measured at 78 K,[20] and we investigate the relation between frustration and the trend of lattice distortion. To probe the details of the long-range exchange interaction, we expanded the parent tetragonal cell to a 2x2x2 supercell, containing 16 Fe and 16 Os sites each and a total of 160 atoms in the cell. We considered ten independent exchange pathways connecting various Fe and Os sites, as illustrated in Fig.3. Some of the simplest representative spin configurations and the corresponding energy expressions are listed in Table1.

*Table1. Four simplest representative magnetic configurations of the Fe and Os ions for the different states used to determine the magnetic interactions. The energy expressions for the specified magnetic configurations are given below the table. ↑ and ↓ represent the two opposite spin directions at the magnetic sites.*

| Site | | Fe | | | | | | | | | | | | | | | | Os | | | | | | | | | | | | | | | |
|---|---|---|---|---|---|---|---|---|---|---|---|---|---|---|---|---|---|---|---|---|---|---|---|---|---|---|---|---|---|---|---|---|---|
| Site No. | | 1 | 2 | 3 | 4 | 5 | 6 | 7 | 8 | 9 | 10 | 11 | 12 | 13 | 14 | 15 | 16 | 1 | 2 | 3 | 4 | 5 | 6 | 7 | 8 | 9 | 10 | 11 | 12 | 13 | 14 | 15 | 16 |
| FM | E1 | ↑ | ↑ | ↑ | ↑ | ↑ | ↑ | ↑ | ↑ | ↑ | ↑ | ↑ | ↑ | ↑ | ↑ | ↑ | ↑ | ↑ | ↑ | ↑ | ↑ | ↑ | ↑ | ↑ | ↑ | ↑ | ↑ | ↑ | ↑ | ↑ | ↑ | ↑ | ↑ |
| AFM1 | E2 | ↑ | ↑ | ↑ | ↑ | ↑ | ↑ | ↑ | ↑ | ↑ | ↑ | ↑ | ↑ | ↑ | ↑ | ↑ | ↑ | ↓ | ↓ | ↓ | ↓ | ↓ | ↓ | ↓ | ↓ | ↓ | ↓ | ↓ | ↓ | ↓ | ↓ | ↓ | ↓ |
| AFM2 | E3 | ↑ | ↑ | ↑ | ↑ | ↑ | ↑ | ↑ | ↑ | ↓ | ↓ | ↓ | ↓ | ↓ | ↓ | ↓ | ↓ | ↓ | ↓ | ↓ | ↓ | ↓ | ↓ | ↓ | ↓ | ↑ | ↑ | ↑ | ↑ | ↑ | ↑ | ↑ | ↑ |
| AFM3 | E4 | ↑ | ↑ | ↑ | ↑ | ↑ | ↑ | ↑ | ↑ | ↓ | ↓ | ↓ | ↓ | ↓ | ↓ | ↓ | ↓ | ↑ | ↑ | ↑ | ↑ | ↑ | ↑ | ↑ | ↑ | ↓ | ↓ | ↓ | ↓ | ↓ | ↓ | ↓ | ↓ |

$$E_1^{FM} = 96J_1 + 48J_2 + 32J_3 + 64J_4 + 32J_5 + 16J_6 + 72J'_3 + 144J'_4 + 72J'_5 + 36J'_6$$

$$E_2 = -96J_1 - 48J_2 + 32J_3 + 64J_4 + 32J_5 + 16J_6 + 72J'_3 + 144J'_4 + 72J'_5 + 36J'_6$$

$$E_3 = 96J_1 - 48J_2 + 32J_3 - 64J_4 + 32J_5 + 16J_6 + 72J'_3 - 144J'_4 + 72J'_5 + 36J'_6$$

$$E_4 = -96J_1 + 48J_2 + 32J_3 - 64J_4 + 32J_5 + 16J_6 + 72J'_3 - 144J'_4 + 72J'_5 + 36J'_6$$

Similarly we constructed the linear algebraic equations for all other spin configurations, using the same methodology described in Table 1. We solved these simultaneous coupled linear algebraic equations to extract the relevant J values. Figure. 3 shows various dominant exchange interaction paths connecting different Fe and Os sites, and Table 2 provides a list of dominant magnetic interactions. A simple trend can be identified, the out-of-plane interactions are much larger than their in-plane counterparts. For example, the in-plane Fe-Os nearest neighbor interaction $J_1$ is very small, whereas the out-of-plane Fe-Os nearest neighbor interaction $J_2$ is significantly stronger. All Os-Os interactions (from $J_3$ to $J_6$) are much stronger than their counterpart Fe-Fe interactions (from $J'_3$ to $J'_6$), which can be attributed to the fact that the Os-5*d* states are more extended than the Fe-3*d* states. Interestingly, both Fe-Fe and Os-Os nearest neighbor interactions ($J_3$ and $J'_3$) are much smaller than the next neighbor interactions ($J_5$, $J_6$ for Os-Fe-Os and $J'_5$, $J'_6$ for Fe-Os-Fe ). The counter intuitive trend of smallest nearest neighbor Fe-Os magnetic interactions ($J_1$-$J_2$) can be understood from the energy level position of Fe and Os states in the DOS. The larger hoppings are expected to occur between Fe-$t_{2g}$ and Os-$t_{2g}$ states compared to those between Fe-$e_g$ and Os-$t_{2g}$. In the DOS, Fe-$t_{2g}$ and Os-$t_{2g}$ have hardly any overlap in the

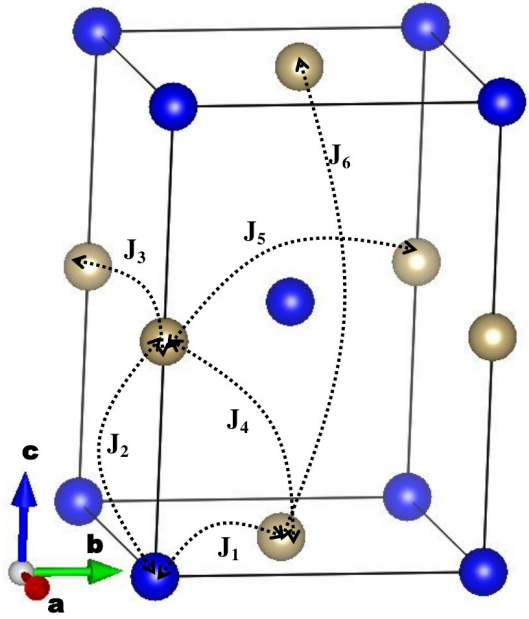

*Fig.3. Different magnetic exchange paths connecting different Fe (blue balls ) and Os (brown balls) sites in the tetragonal unit cell of $Sr_2FeOsO_6$. Sr and O atoms are omitted from the structure figure for clarity. $J_1$ and $J_2$ are interaction paths between Fe-Os while Os-Os interactions are denoted by $J_3$ to $J_6$ . The Fe-Fe counterparts ( with same nomenclature as that of Os-Os) $J'_3$ to $J'_6$ are not marked in the figure for sake of simplicity.*

*Table2. Calculated magnetic exchange interactions for different paths, as shown in Fig. (3). The values are listed in the table below.*

|  | **Interaction Paths** | **Values (meV)** | **Type** |
|---|---|---|---|
| $J_1$ | Fe-Os (in plane) | 0.1 | AFM |
| $J_2$ | Fe-Os (out of plane) | 4.2 | FM |
| $J_3$ | Os-Os (in plane) | 0.2 | FM |
| $J_4$ | Os-Os (out of plane) | 3.3 | AFM |
| $J_5$ | Os-Os (in plane) | 6.8 | FM |
| $J_6$ | Os-Os (out of plane) | 12.8 | AFM |
| $J'_3$ | Fe-Fe (in plane) | 0.1 | FM |
| $J'_4$ | Fe-Fe (out of plane) | 2.2 | AFM |
| $J'_5$ | Fe-Fe (in plane) | 1.3 | FM |
| $J'_6$ | Fe-Fe (out of plane) | 2.2 | FM |

energy scale, whereas Fe-$e_g$ and Os-$t_{2g}$ states are substantially overlapped. Therefore the interaction between Fe and Os is expected to be small. The fact that next-nearest neighbor exchange interactions are stronger than the nearest neighbor interactions has been discussed in the literature in the context of 3$d$-5$d$ double perovskite systems.[31,32] These results can be visualized prominently using localized Wannier functions representations. Figure 4 shows plots of effective orbitals corresponding to the downfolded Wannier like orbitals located at Os and Fe sites. The central part is comprised of Os- or Fe- $d$ characters, while the tails situated at the different sites are shaped according to integrated out orbitals and the weight of the neighboring tail dictates the strength of the interaction between the sites. Figure 4a shows that the $J_2$ (Os-Fe) interaction is very small compared with the $J_6$ (Os-Fe-Os) interaction, as there are large *d* tails at the connecting Os site compared to the connecting Fe site. Similarly Fig. 4b shows that the in-plane nearest neighbor $J_1$ (Os-Fe) interaction is very small (no *d* tails at the connecting Os site) compared with the in-plane next-nearest neighbor $J'_5$ (Fe-Os-Fe) interaction. The out-of-plane next nearest neighbor (Fe-Os-Fe) $J'_6$ interaction is substantially smaller, than the out-of-plane $J_6$ (Os-Fe-Os) interaction (Fig. 4a). Additionally Fe-Fe and Os-Os are coupled FMs in the *ab* plane and coupled AFM along the crystallographic elongated *c* axis.

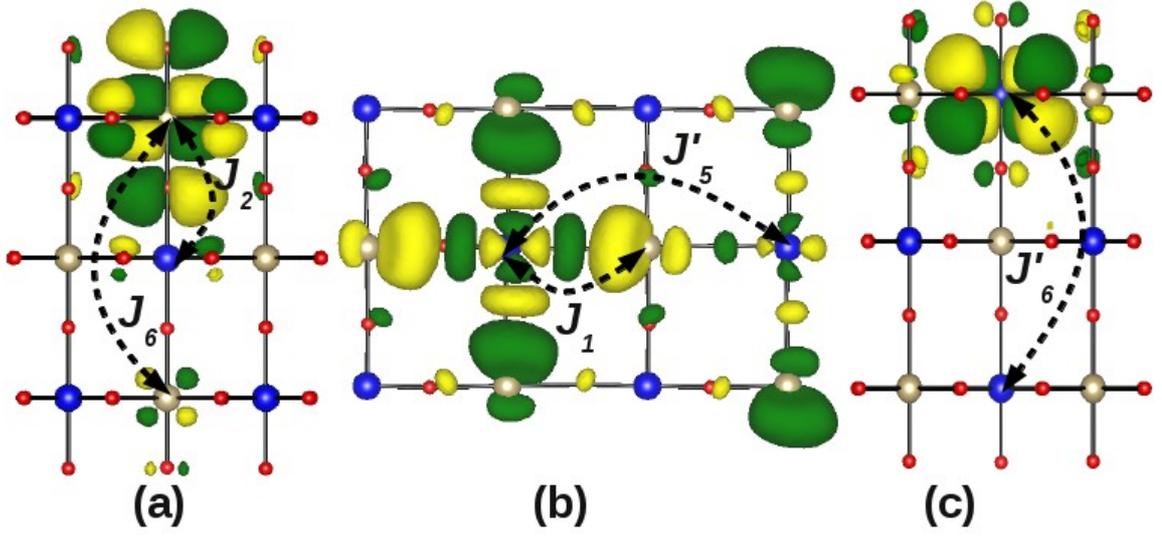

*Fig.4. The Wannier functions of (a) Os-5$d_{yz}$, (b) Fe-3$d_{x2-y2}$ and (c) Fe-3$d_{yz}$ orbitals centered at Os (yellow balls) and Fe sites (blue balls). The exchange interactions $J_2$ -$J_6$ , $J_1$-$J'_5$ and $J'_6$ are shown in (a), (b) and (c) respectively. Green and yellow colors represent surfaces with isovalues +0.2 and -0.2, respectively. Red balls represent O atoms.*

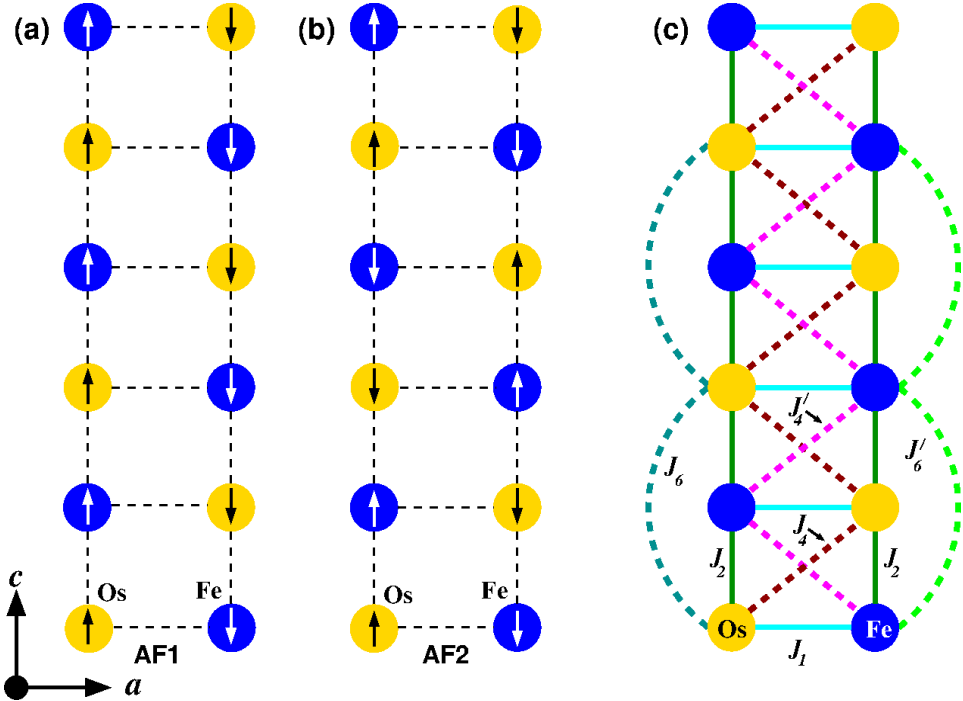

*Fig.5: Schematic spin lattice model of an exclusively Fe-Os sub lattice. (a) and (b) show Fe-Os spin arrangement for AF1 and AF2 phases, respectively. (c) represents the Fe-Os effective spin lattice model with $J_1$ -$J_2$ -$J_4$ -$J'_4$ -$J_6$ -$J'_6$ interaction paths. This figure shows the competition between $J_2$ and $J_6$ which leads to the "↑↑↓↓" spin arrangement in the AF2 phase.*

In the next step we will try to explain the observed "↑↑↓↓↑↑↓↓" spin arrangements between Os and Fe sites (Os-Fe-Os-Fe-Os-Fe-Os-Fe) in the AF2 phase and the magnetic transition from the AF1 phase to the AF2 phase, on the basis of our calculated magnetic exchange interactions. The main objective of the present study is to determine the proper reasoning to explain the magnetic transition from the AF1 phase to the AF2 phase. As an example, we illustrate the simple effective spin model in Fig. 5. All exchange interactions (except $J_6$) are fully satisfied in the AF1 phase (as shown in Fig.5a), and these interactions reflect the experimentally observed phenomena,[21] in which FM type Fe-Os chains form along the *c*-axis, while these chains couple as AFM in the *ab* plane. However, $J_6$ induces an AFM interaction along the Fe-Os chain, and this interaction attains the largest amplitude among all J values. Clearly, the existence of the AFM type $J_6$ violates all other out-of-plane interactions (including $J_2$, $J_4$, $J'_4$ and $J'_6$), and causes a significant frustration effect. Along the Fe-Os chain in the direction of the *c*-axis, the AFM type $J_6$ favors that the two Os sites involved show anti-parallel spins, while the FM type $J_2$, (the second largest J value) requires all Os spins to be parallel (as shown in Fig. 5c). When the temperature is sufficiently low (below the second transition temperature of 67 K), the strongest $J_6$ overtakes the competition and realizes AFM type Os-Os coupling (as shown in Fig.5b), which induces the phase transition from AF1 (Fig. 5a) to AF2 (Fig. 5b). Consequently in the AF2 phase, the compound distorts, with alternate elongation and contraction of the Fe-Os distances along the *c*-axis, and this distortion occurs in order to compensate for the unsatisfied $J_2$ interaction. For simplicity, this compound can be regarded as a one-dimensional Ising model with an effective Hamiltonian of the form, $H = J_2 \sum_{FeOs} \vec{S}_{Fe} \cdot \vec{S}_{Os} + J_6 \sum_{OsOs} \vec{S}_{Os} \cdot \vec{S}_{Os}$, where the nearest-neighbor interaction ($J_2$) and next nearest neighbor interaction ($J_6$) compete with each other.

## IV. CONCLUSION

In summary, we studied the ordered tetragonal 3*d*-5*d* double perovskite $Sr_2FeOsO_6$ using *ab-initio* calculations. We calculated the effective magnetic exchange interactions between different Fe and Os sites, and these calculations reveal the importance of long range super-super exchange interactions in this system. On the basis of our calculated exchange interactions, we analyzed the AF1 to AF2 magnetic phase transition. The strong frustration effect between nearest neighbor Fe-Os exchange interactions and next nearest neighbor Os-Os exchange interactions along the *c*-axis is found to cause the magnetic phase transition from AF1 to AF2, which results in lattice distortions.

## V. ACKNOWLEDGEMENT

We are grateful to P. Adler for fruitful discussion. This work is financially supported by the ERC Advanced Grant (291472).

---